\documentclass[A4paper,twocolumn]{article}
\usepackage{amssymb}

\usepackage{graphicx}
\usepackage{amsmath}

\newtheorem{lemma}{Lemma}
\newtheorem{theorem}{Theorem}

\begin{document}

\title{Necessary and sufficient criterion for k-separability of N-qubit noisy GHZ
states}
\author{Xiao-yu Chen$^a$\thanks{%
Email:xychen@zjgsu.edu.cn}, Li-zhen Jiang$^a$ Zhu-an Xu$^b$ \\
{\small {$^a$College of Information and Electronic Engineering, Zhejiang
Gongshang University, Hangzhou, Zhejiang 310018, China }}\\
{\small {$^b$Department of Physics, Zhejiang University, Hangzhou, Zhejiang
310027, China }}}
\date{}
\maketitle

\begin{abstract}
A Multipartite entangled state has many different kinds of entanglement
specified by the number of partitions. The most essential example of
multipartite entanglement is the entanglement of multi-qubit
Greenberger-Horne-Zeilinger (GHZ) state in white noise. We explicitly
construct the entanglement witnesses for these states with stabilizer
generators of the GHZ states. For a $N$ qubit GHZ state in white noise, we
demonstrate the necessary and sufficient criterion of separability when it
is divided into $k$ parties with $N\leq 2k-1$ for arbitrary $N$ and $k$. The
criterion covers more than a half of all kinds of partial entanglement for $%
N $-qubit GHZ states in white noise. For the rest of multipartite
entanglement problems, we present a method to obtain the
sufficient conditions of separability. As an application, we
consider $N$ qubit GHZ state as a codeword of the degenerate
quantum code passing through depolarizing channel. We find that
the output state is neither genuinely entangled nor fully
separable when the quantum channel capacity reduces from positive to zero.

PACS number(s): 03.67.Mn; 03.65.Ud

\end{abstract}

\section{Introduction}

Entanglement in multipartite systems is a key resource for quantum information and communication protocols\cite{Horodecki,Guhne}.  In experiments, various multipartite entangled states have been generated\cite{Blatt,Pan,Neumann,Sackett,Zhao,Gao}.  The question whether or not an experimentally generated multipartite state is (partially) entangled has become a highly relevant topic for quantum information theory.  Although an enormous amount of work has been devoted to detect multipartite entanglement \cite{Seevinck08,Huber,Seevinck,Badziag,Hason, Laskowski,Vicente,Ananth, GuhnePLA,Chen2015,Chen2017,Guhne2011}, we are still very far away from the characterization of multipartite entanglement.
The experimental detection of
entanglement is typically done via the construction of proper entanglement
witnesses \cite{Guhne,Chen2017,Terhal}, and multipartite witnesses have been
considered \cite{Loock,Bourennane,Eisert,Chruscinski}. 

Among all the multipartite entangled states, the Greenberger-Horne-Zeilinger (GHZ) states are  the simplest ones. The GHZ entanglement, originally introduced to explore the extreme violation
of local realism against quantum mechanics, is an important resource for
multipartite quantum communication tasks such as quantum cryptographic
conferencing (QCC) \cite{Shibata,Yin,Ma}, quantum secret
sharing (QSS) \cite{Bose,Cleve,Tittel,Bell}and
third-man quantum cryptography \cite{Zukowski}. Typically, the experimental
preparations of multipartite entangled states \cite{Blatt,Pan,Neumann} are the $N$ qubit GHZ states. The imperfection and noise in the
preparations are usually described by white noise. Most probably, an
experimental prepared multipartite state is the mixture of $N$ qubit GHZ
state with white noise (noisy GHZ state, also known as generalized Werner state \cite{Pittenger} or Werner-Popescu state \cite{Abe,Nayak}). Meanwhile
the noisy GHZ state describes the output state of the GHZ state passing through a depolarizing channel.

Suppose there is a composed Hilbert space $\mathcal{H}=\mathcal{H}_{1}\otimes\cdot\cdot\cdot\otimes\mathcal{H}_{n}$.
Consider a partition $\mathcal{I}=\{\mathcal{I}_{1},..., \mathcal{I}_{k}\}$ of the index set $\mathcal{J} = \{1, . . . , n\}$. A quantum state $\sigma_{\mathcal{I}}$ is called separable for the given partition $\mathcal{I}$, if it can be written as a classical mixture of product states:
\begin{equation}\label{1a}
\sigma_{\mathcal{I}}=\sum_{i}q_{i}|\psi^{(i)}_{\mathcal{I}_{1}}\rangle\langle\psi^{(i)}_{\mathcal{I}_{1}}|\otimes\cdot\cdot\cdot\otimes|\psi^{(i)}_{\mathcal{I}_{k}}\rangle\langle\psi^{(i)}_{\mathcal{I}_{k}}|,
\end{equation}
with $q_{i}$ being a classical probability distribution, $|\psi^{(i)}_{\mathcal{I}_{j}}\rangle$ is a pure state of subset $\mathcal{I}_{j}$.  A state $\sigma$ is called $k$-separable if it can be written to be
\begin{equation}\label{1b}
 \sigma=\sum_{\mathcal{I}:|\mathcal{I}|=k}q'_{\mathcal{I}}\sigma_{\mathcal{I}}.
\end{equation}
Where $|\mathcal{I}|$ is the number of elements in the set $\mathcal{I}$, $q'_{\mathcal{I}}$ is a classical probability distribution. The summation is over all possible $k$ partite partitions. If a quantum state cannot be written in the form of Eq.(\ref{1b}), it is referred to as $k$-inseparable. A $2$-inseparable (not biseparable) state is also called genuinely entangled. A $N$-inseparable (not fully separable) state is an entangled state.

Although the biseparable $(k=2)$ condition and the fully separable $(k=N)$ condition
for a $N$ qubit noisy GHZ state are known \cite{Seevinck,Cirac,Schack}, the conditions for the $k$-separability with $2<k<N$ almost remain to be uncovered for the state.

The partition of an $N$ qubit noisy GHZ state is greatly simplified by the
symmetry of the state. In fact we just count the number of qubits in a party
without considering which qubits are in the party. Let the number of qubits in subset $\mathcal{I}_{j}$ be $n_j$. We may alternatively denote the partition as  $n_1|n_2|n_3|\ldots |n_j|\ldots |n_k$. Then $\sum_{j=1}^kn_j=N$.

We will show that a properly chosen entanglement witness constructed with stabilizer generators of GHZ state can detect the
$k$-separability of a noisy GHZ state.  The paper is organized as follows: In
section 2, we describe the entanglement witnesses for noisy GHZ states. In
section 3 and section 4, the necessary and sufficient conditions are
demonstrated for the $k$ separability of $N$ qubit noisy GHZ states when $%
N\leq 2k-1$. We also give the witnesses explicitly in section 3. In section
5, the sufficient conditions of all the other separabilities for $N\leq 12$
are shown. In section 6, we discuss the application of our findings to GHZ
state in depolarizing channel. The conclusions are drown in section 7.
Several lemmas or theorem are proven in the appendix.

\section{Entanglement Witnesses of Noisy GHZ states}

A witness of multipartite entanglement (more strictly $k$-inseparability) is a Hermitian
operator $\hat{W}$ with $tr(\rho_{s} \hat{W})\geq 0$ for all $k$-separable state $\rho_{s} $, and $tr(\rho \hat{W})<0$ for at least one state $\rho $. A witness $\hat{W}$ is optimal if $tr(\rho_{s} \hat{W})= 0$ is achieved for some $k$-separable state $\rho_{s}$. Let $\hat{M}$ be a Hermitian operator, the optimal witness can be produced as
 $\hat{W}=\Lambda\Bbb{I}-\hat{M},$
where $\Bbb{I}$ is the identity operator of the quantum system, and
\begin{equation}
\Lambda=\max_{\rho _s}tr(\rho _s\hat{M}),  \label{wee1}
\end{equation}
The method leading to a k-separable criterion is to first choose an Hermitian operator $\hat{M}$, then find the $\Lambda$ with respect to all the $k$-separable state $\rho _s$. If there is a state $\rho$ with $tr(\rho\hat{M})>\Lambda$, then the state $\rho$ is called $k$-inseparable. Let $\mathcal{L}=\frac{\Lambda}{tr(\rho\hat{M})}$, the necessary criterion of $k$-separability is
\begin{equation}\label{wee1a}
   \mathcal{L}\leq 1.
\end{equation}
 The criterion may not be efficient by the ad hoc nature of choosing $\hat{M}$. For a give state $\rho$, we may modify $\hat{M}$ to decrease $\mathcal{L}$ such that the criterion is more efficient. For a noisy $N$ qubit GHZ state, we propose a proper $\hat{M}$ then prove the necessary criterion to be also sufficient.

A noisy $N$ qubit GHZ state is
\begin{equation}
\rho _{ghzN}=p\left| GHZ_N\right\rangle \left\langle GHZ_N\right|
+\frac{(1-p)}{2^{N}}\Bbb{I}.  \label{wee2}
\end{equation}
Where $\left| GHZ_N\right\rangle =\frac 1{\sqrt{2}}(\left| 0\right\rangle
^{\otimes N}+\left| 1\right\rangle ^{\otimes N})$. In the following, we will denote $\left|
\chi \right\rangle \left\langle \chi \right| $ as $\left| \chi \right\rangle
\left\langle \cdot \right| $ for short. The stabilizer of $\left|
GHZ_N\right\rangle $ is an operator Abel group created by generators $\{%
\hat{K}_1,\ldots ,\hat{K}_N\}$ such that $\hat{K}\left|
GHZ_N\right\rangle =\left| GHZ_N\right\rangle $ for any stabilizer operator $\hat{K}=%
\hat{K}_1^{j_1}\cdots \hat{K}_N^{j_N}$ with $j_i=0,1.$ The
generators are
\begin{eqnarray}
\hat{K}_1 &=&\sigma _1\otimes \sigma _1\otimes \cdots \sigma _1,
\nonumber \\
\hat{K}_2 &=&\sigma _3\otimes \sigma _3\otimes I\otimes \cdots I,
\nonumber \\
\hat{K}_3 &=&\sigma _3\otimes I\otimes \sigma _3\otimes I\otimes \cdots
I,  \nonumber \\
&&\ldots ,  \nonumber \\
\hat{K}_N &=&\sigma _3\otimes I\otimes \cdots I\otimes \sigma _3,
\label{ws1}
\end{eqnarray}
with $\sigma _1,\sigma _2,\sigma _3\ $ being Pauli matrices and $I=\sigma _0$
is the $2\times 2$ identity matrix. The projector onto the GHZ state has a direct representation in terms of the
corresponding stabilizer, namely,

\begin{equation}
\left| GHZ_N\right\rangle \left\langle GHZ_N\right| =\frac 1{2^N}\sum_{%
\hat{K}\in stabilizer}\hat{K}  \label{ws2}
\end{equation}
We have
\begin{equation}\label{wq1}
tr(\rho _{ghzN}\hat{K})=\left\{
\begin{array}{l}
1,\text{ for }\hat{K}=\Bbb{I} \\
p,\text{ otherwise.}
\end{array}
\right.
\end{equation}
for a stabilizer operator $\hat{K}$.

Since $\rho_{ghzN}$ is a linear mixture of stabilizer operators, we may
suppose the Hermitian operator $\hat{M}$ to be a linear combination of
stabilizer operators instead of all elements of Pauli group. The most
general form of a Hermitian operator $\hat{M}$ is
\begin{equation}\label{wq2}
  \hat{M}=\sum_{j_1,\ldots j_N=0}^1M_{j_1j_2\ldots j_N}\hat{K}%
_1^{j_1}\cdots \hat{K}_N^{j_N}.
\end{equation}
An observation of (\ref{ws1}) shows that the generators $K_{2},...,{K_{n}}$ constitute a subgroup of the stabilizer. The subgroup is responsible for the diagonal entries of $\rho_{ghzN}$. The coset (the product of $K_{1}$ with the subgroup) is responsible for the anti-diagonal entries of $\rho_{ghzN}$. Due to the symmetry of $\rho_{ghzN}$, it is reasonable that the Hermitian operator (thus the witness) also possesses the symmetry. We then assume $M_{j_1j_2\ldots j_N}$ to be a function of $|\mathbf{j}|$ instead of $\mathbf{j}$, where $\mathbf{j}=(j_{2},...,j_{N})$. Furthermore, we assume $\hat{M}=\hat{M}_{D}+\hat{M}_{A},$
with
\begin{eqnarray}
\hat{M}_{D} &=&\sum_{\mathbf{j}}M_{\left\lceil \left| \mathbf{j}\right|
/2\right\rceil }\hat{K}_2^{j_2}\cdots \hat{K}_N^{j_N},  \label{wee3}
\\
\hat{M}_{A} &=&\hat{K}_1\sum_{\mathbf{j}}\hat{K}_2^{j_2}\cdots
\hat{K}_N^{j_N},  \label{wee3b}
\end{eqnarray}
The reasons are as follows. All coefficients $M_{1j_2\ldots j_N}$ in $\hat{M}_{A}$ are set to be equal (without loss of generality, they are set to be $1$), since in
computational basis we have $\hat{M}_{A}=2^{N-1}(\left| 0\right\rangle
^{\otimes N}\left\langle 1\right| ^{\otimes N}+\left| 1\right\rangle
^{\otimes N}\left\langle 0\right| ^{\otimes N})$ due to (\ref{ws1}) and (\ref
{ws2}). Such a choice of $\hat{M}_{A}$ is reasonable since it is
proportional to the anti-diagonal part of the state $\rho _{ghzN}$. All the stabilizer operators
in $\hat{M}_{D}$ contribute to diagonal part of the Hermitian operator $\hat{M}$ (thus the witness operator) in computational basis. The
coefficients $M_{\left\lceil \left| \mathbf{j}\right| /2\right\rceil }$ in $\hat{M}_{D}$ are so chosen such that the stabilizer operators with equal number of $\sigma
_3 $ have the same coefficient. For example, all the generators $K_i$ $%
(i=2,...,N)$ have two $\sigma _3$ in their tensor product expressions, the
coefficient for these stabilizer operators is $M_1$. The stabilizer operators with two generators $K_iK_j$
$(i,j=2,...,N;$ $i\neq j)$ also have two $\sigma _3$ in their tensor product
expressions, the coefficient for them is $M_1$ too. The coefficients
in $\hat{M}_{D}$ are so chosen that we keep the qubit exchange symmetry
for the Hermitian operator. We also assume $M_0=0$ since a nonzero $M_0$ only leads to a displacement of $\Lambda$.

Notice that $tr(\rho _{ghzN}\hat{M}_{A})=2^{N-1}p$. There are $C_N^{2i}=\frac{N!}{(N-2i)!(2i)!}$ different tensor products $\sigma_{3}^{j_{1}}\otimes...\otimes\sigma_{3}^{j_{N}}$ subject to $ \sum_{l} j_{l}=2i,  j_{l}\in\{0,1\}$, each of the tensor products is a stabilizer operator.
We then have
\begin{equation}
tr(\rho _{ghzN}\hat{M})=p(\sum_{i=1}^{\left\lfloor N/2\right\rfloor
}M_iC_N^{2i}+2^{N-1}),  \label{wee4}
\end{equation}
From (\ref{wee1a}), the necessary condition for the $k$-separability of $\rho
_{ghzN}$ then is
\begin{equation}
p\leq \frac{\Lambda}{(\sum_{i=1}^{\left\lfloor N/2\right\rfloor
}M_iC_N^{2i}+2^{N-1})}.  \label{wee5}
\end{equation}
In convention, $\Lambda$ is with respect to $k$-partition.

\section{Necessary condition of $k$ ($k\geq \frac{N+1}2$) separability}

A $k$-partite partition $n_1|n_2|n_3|\ldots |n_l|\ldots |n_k$ splits the $N$
qubit system into $k$ parties. There are many $k$-partite partitions since
the number of qubits in each party varies from $1$ to at most $N-k+1.$ The notation of the partition can be shorten as $n_1^{2}|n_3|\ldots |n_l|\ldots |n_k$ if $n_{2}=n_{1}$. So alternatively, we may denote the $k$-partite partition as
 $1^{N_1}|2^{N_2}|3^{N_3}|...|m^{N_m}$ with $%
k=\sum_{i=1}^mN_i,$ $N=\sum_{i=1}^miN_i$. The number of $i$ qubit parties is
$N_i$ in the partition. One of the main findings of this paper is that the coefficients $M_{i}$ of the Hermitian operator $M$ should form an arithmetic progression.
 \begin{equation}
M_i=\frac{4i-N}{N_{1}},  \label{wee34}
\end{equation}
We stress that $N_{1}$ is the number of single qubit parties in the partition. We start with a special $k$-partite partition
 $1^{k-1}|(N-k+1)$ for the necessary condition of $k$-separability.

\subsection{Partition $1|1|\ldots |1|L$}

To obtain $\Lambda$, the maximal mean of the Hermitian operator $\hat{M}$ with respect to all $k$-separable states, we only need to consider
pure $k$-separable states. For the partition $1^{N_1}|L$, we have $N_{1}=k-1$, $L=N-k+1$, there is a $
k$-separable pure state
\begin{equation}\label{wee34a}
\rho _s=\bigotimes_{i=1}^{N_1}\varrho _i\otimes \varrho _{N_1+1,\ldots ,N},
\end{equation}
where $\varrho _i=\frac 12(I+x_i\sigma _1+y_i\sigma _2+z_i\sigma _3),$
$(i=1,\ldots ,N_1)$ is the pure state of the $ith$ qubit, with $%
x_i^2+y_i^2+z_i^2=1;$ $\varrho _{N_1+1,\ldots ,N}\equiv \left| \psi
\right\rangle \left\langle \psi \right| $ is the pure state for the last $%
L$ qubits. Then
\begin{eqnarray}
tr(\rho _s\hat{M}_{D}) &=&\sum_{\mathbf{j}}M_{\left\lceil \left| \mathbf{j}%
\right| /2\right\rceil }z_1^{j^{\prime }}\prod_{i=2}^{N_1}z_i^{j_i}
\nonumber \\
&&\times \left\langle \psi \right| \bigotimes_{l=N_1+1}^N\sigma
_3^{j_l}\left| \psi \right\rangle ,  \label{wqq1}
\end{eqnarray}
where $j^{\prime }=$mod$(\left| \mathbf{j}\right| ,2).$ We also have
\begin{eqnarray}
tr(\rho _s\hat{M}_{A}) &=&\sum_{\mathbf{j}}(-i)^{\left| \mathbf{j}\right|
+j^{\prime }}x_1^{1-j^{\prime }}y_1^{j^{\prime
}}\prod_{i=2}^{N_1}x_i^{1-j_i}y_i^{j_i}  \nonumber \\
&&\left\langle \psi \right| \bigotimes_{l=N_1+1}^N\sigma _1^{1-j_l}\sigma
_2^{j_l}\left| \psi \right\rangle .  \label{wqq2}
\end{eqnarray}
We may write $tr(\rho _s\hat{M})=\left\langle \psi \right| \mathcal{M}%
\left| \psi \right\rangle ,$ where $\mathcal{M}$ is a $2^{L}\times
2^{L}$ matrix. The maximum of $tr(\rho _s\hat{M})$ then is equal to
the maximal eigenvalue of $\mathcal{M}$. Denote the diagonal elements of
matrix $\mathcal{M}$ as $m_{\mathbf{i,i}}$, where $\mathbf{i=(}%
i_1,i_2,\ldots ,i_{L})$ is a binary string and we denote the weight of $%
\mathbf{i}$\textbf{\ }as $\left| \mathbf{i}\right| $. Then $m_{%
\mathbf{i,i}}=\Gamma _{\left| \mathbf{i}\right| }$ with $\Gamma _{\left|
\mathbf{i}\right| }$ defined in (\ref{wee29a}) of the appendix. Let $\mathcal{M}_{\mathbf{i}
}=m_{\mathbf{i,i}}|\mathbf{i}\rangle\langle\mathbf{i}|$,
then we have

\begin{lemma}
All the non-diagonal entries of matrix $\mathcal{M}$ is nullified except the
entries of $\left| 0^{\otimes L}\right\rangle \left\langle 1^{\otimes
L}\right| $ and $\left| 1^{\otimes L}\right\rangle \left\langle
0^{\otimes L}\right| $, namely, $\mathcal{M}$ is block diagonalized as $%
\mathcal{M=M}_0\bigoplus_{\mathbf{i\neq 0,i\neq 1}}\mathcal{M}_{\mathbf{i}%
},$ where $\mathbf{0=(}0,\ldots ,0),\mathbf{1=(}1,\ldots ,1).$ The submatrix
$\mathcal{M}_0$ in the computational basis $\left| 0^{\otimes
L}\right\rangle $and $\left| 1^{\otimes L}\right\rangle $ is
\begin{equation}
\mathcal{M}_0=\left[
\begin{array}{ll}
\Gamma _0 & 2^{L-1}(c-id) \\
2^{L-1}(c+id) & \Gamma _{L}
\end{array}
\right]  \label{wee35}
\end{equation}
where $c=\prod_{i=1}^{N_1}\sin \theta _i\cos \varphi ,$ $d=\prod_{i=1}^{N_1}%
\sin \theta _i\sin \varphi $, with $\varphi =\sum_{j=1}^{N_1}\varphi _j.$
\end{lemma}
Moreover, we have
\begin{lemma}
The maximal eigenvalue of $\mathcal{M}_0$ is
\begin{equation}
\Lambda_{0}=\frac N{N_{1}}.  \label{wee36}
\end{equation}
\end{lemma}
Using (\ref{wee34}), the necessary criterion (\ref{wee5}) is simplified to
\begin{equation}
p\leq \frac{\Lambda}{\frac N{N_{1}}+2^{N-1}}.  \label{wee34a}
\end{equation}
\begin{theorem}
The maximal eigenvalue of $\mathcal{M}$ is $\Lambda=\frac{N}{N_{1}}$.
The necessary condition of separability for $\rho _{ghzN}$ is
\begin{equation}
p\leq [1+2^{N-1}N_{1}/N]^{-1}  \label{wee40}
\end{equation}
with respect to the partition $1^{N_1}|L.$
\end{theorem}

The proofs of Lemma 1, Lemma 2 and Theorem 1 are shown in the appendix.

\subsection{Further splits of the last $L$ qubits}

We consider further splits of the last $L$ qubits into two parties. Denote the partition after split as $1^{N_1}|\overline{L}=1^{N_1}|l|L-l$ with $N=N_{1}+L, k=N_{1}+2$. Keep in mind that we have exhausted all the single qubit parties
into $1^{N_1}$, the split of the last $L$ qubits do not produce new
single qubit partite. The smallest piece from the further split of the last $%
L$ qubits is a two qubit partite, namely $l\geq 2,  L-l\geq 2$. On the other hand, we have noticed that the maximum of $tr(\rho _s\hat{M})$ do not increase by the split of
the last $L$ qubits. This is due to the fact that when we split the pure
state of the last $L$ qubits $\left| \psi \right\rangle $ into a product
of pure states, the matrix $\mathcal{M}$ does not change since we only
change the last part of $\rho _s$ and do not change the state of the first $%
N_1$ qubits and the operator $\hat{M}$. The maximal eigenvalue of $%
\mathcal{M}$ is achieved when $\left| \psi \right\rangle $ is the
corresponding eigenvector, namely $\left\langle \psi \right| \mathcal{M}%
\left| \psi \right\rangle \geq \left\langle \psi ^{\prime }\right| \mathcal{M%
}\left| \psi ^{\prime }\right\rangle $ for all the other state $\left| \psi
^{\prime }\right\rangle $ including the product state produced by the split
of the last $L$ qubits. We conclude that the maximum of $tr(\rho _s%
\hat{M})$ for partition $1^{N_1}|\overline{L}$ can not exceed $%
\frac N{N_1}.$

Next we should prove that the maximal eigenvalue $\frac N{N_1}$ of $\mathcal{M}$ is achievable by the state with partition $1^{N_1}|\overline{L}$. Let $|\psi^{\prime}\rangle=|\psi_{l}\rangle|\psi_{L-l}\rangle$ be the product state of the partition $\overline{L}$, with rather generic states $|\psi_{l}\rangle=\alpha_{1}|0^{\otimes l}\rangle+\beta_{1}|1^{\otimes l}\rangle$ and $|\psi_{L-l}\rangle=\alpha_{2}|0^{\otimes L-l}\rangle+\beta_{2}|1^{\otimes L-l}\rangle$ subject to the constrain that $\alpha_{1}\alpha_{2}|0^{\otimes L}\rangle+\beta_{1}\beta_{2}|1^{\otimes L}\rangle\equiv |\psi_{a}\rangle\sim |\psi\rangle $ is the unnormalized eigenvector corresponding to the largest eigenvalue of $\mathcal{M}$. Let $|\psi_{b}\rangle=\alpha_{1}\beta_{2}|0^{\otimes l}\rangle|1^{\otimes l-l}\rangle+\beta_{1}\alpha_{2}|1^{\otimes l}\rangle|0^{\otimes L-l}\rangle$, then $|\psi'\rangle=|\psi_{a}\rangle+|\psi_{b}\rangle$. Then we have
\begin{eqnarray}
  &\langle\psi'|\mathcal{M}|\psi'\rangle= \langle\psi'|\mathcal{M}_0\oplus\mathcal{M}_{0^{l}1^{L-l}}\oplus\mathcal{M}_{1^{l}0^{L-l}}|\psi'\rangle\nonumber\\
   &=\langle\psi_a|\mathcal{M}_0|\psi_{a}\rangle+\langle\psi_b|\mathcal{M}_{0^{l}1^{L-l}}\oplus\mathcal{M}_{1^{l}0^{L-l}}|\psi_{b}\rangle \nonumber\\
   &=(|\alpha_{1}\alpha_{2}|^2+|\beta_{1}\beta_{2}|^2)\Lambda_{0}+|\alpha_{1}\beta_{2}|^2\Gamma_{L-l}+|\alpha_{2}\beta_{1}|^2\Gamma_{l}\nonumber\\
   &=\frac{N}{N_{1}}.
\end{eqnarray}
The last equality comes from Lemma 2 and the fact that $\Gamma_{l}=\frac{N}{N_{1}}$ for $1<l<L$ proven in the appendix.

Based on the proof that the largest eigenvalue of $\mathcal{M}$ is achievable for $\overline{L}$ part of the partition $1^{N_1}|\overline{L}$, it is straightforward to show that the largest eigenvalue of $\mathcal{M}$ is also achievable for the $2^{N_2}|3^{N_3}|...|m^{N_m}$ part of a partition $1^{N_1}|2^{N_2}|3^{N_3}|...|m^{N_m}$. Theorem 1 then is true for any partition $1^{N_1}|2^{N_2}|3^{N_3}|...|m^{N_m}$.

\subsection{$k$-partite separability}

We then consider the $k$ ($k\geq \frac{N+1}2$) partite separability.
The maximum of $tr(\rho _s\hat{M})$ for a $k$-partite
partition $1^{N_1}|2^{N_2}|3^{N_3}|...|m^{N_m}$ (where $k=\sum_{i=1}^mN_i,$ $N=\sum_{i=1}^miN_i$) is equal to $\frac N{N_1}$ as shown in the last subsection. We
will analyze the possible largest $\frac N{N_1}$ for $k$ ($k\geq \frac{N+1}2$%
) partite separability.

We first distribute each partite with one qubit and we remain $N-k$ qubits.
Then we have many strategies to distribute the remained qubits. The best way
to decrease the number of single qubit parties (in order to increase $\frac
N{N_1}$) is to distribute the remained $N-k$ qubits to $N-k$ parties, then
we have $N_1=k-(N-k)$ single qubit parties and $(N-k)$ two qubit parties,
the resultant partition is $1^{(2k-N)}|2^{(N-k)}$. The condition $k\geq
\frac{N+1}2$ guarantees $N_1\geq 1,$ thus there is at least one single qubit
party. The maximum of $tr(\rho _s\hat{M})$ for the partition $%
1^{(2k-N)}|2^{(N-k)}$ is equal to the maximum of $tr(\rho _s%
\hat{M})$ for the partition $1^{(2k-N)}|2(N-k)$ , the later is $\frac
N{2k-N}.$ Hence we have
\begin{equation}
\Lambda\leq \frac N{k-2N}  \label{wee41a}
\end{equation}
for all $k$ ($k\geq \frac{N+1}2$) partite separable noisy GHZ states. The
achievable upper bound $\frac N{k-2N}$ of $\Lambda$ leads to the following theorem.

\begin{theorem}
The necessary condition of the $k$-separability ($k\geq \frac{N+1}2$) of the noisy $N$ qubit GHZ state $\rho _{ghzN}$ is
\begin{equation}
p\leq [1+2^{N-1}(2k-N)/N]^{-1}.  \label{wee41}
\end{equation}
\end{theorem}

\section{Sufficient conditions for $k$ ($k\geq \frac{N+1}2$) separability}

\subsection{Partition $1^{N_1}|L$}

The way of $tr(\rho _s\hat{M})$ achieving its maximum $\Lambda =\frac
N{N_1}$ hints the sufficient condition. We consider the case that the
maximum value of $tr(\rho _s\hat{M})$ is achieved by the separable state
with $z_i=0$ for $i=1,\ldots ,N_1$, hence the pure state of the $ith$ qubit
is $\varrho _i=\frac 12(I+x_i\sigma _1+y_i\sigma _2)$ with $x_i^2+y_i^2=1$.
We may assume $x_i=\cos \varphi _i$, $y_i=\sin \varphi _i$, then $\varrho
_i=\left| \beta _i\right\rangle \left\langle \beta _i\right| ,$ where $%
\left| \beta _i\right\rangle =\frac 1{\sqrt{2}}(\left| 0\right\rangle
+e^{i\varphi _i}\left| 1\right\rangle )$. The state of the last $L$
qubits is the eigenvector of $\mathcal{M}_0$ corresponding to eigenvalue $%
\Lambda _0=\frac{N}{N_{1}}$. The eigenvector is
\begin{equation}
|\psi\rangle =\frac 1{\sqrt{2}}(| 0^{\otimes
L}\rangle +e^{-i\varphi }| 1^{\otimes L}\rangle ).
\label{wee19}
\end{equation}
Where $\varphi=\sum_{i=1}^{N_{1}}\varphi_{i}$ as defined in Lemma 1.  The $k (k=N_{1}+1)$ partite separable pure state is
\begin{equation}
\left| \Omega \right\rangle =\frac 1{\sqrt{2^{N_1}}}\prod_{i=1}^{N_1}(\left|
0\right\rangle +e^{i\varphi _i}\left| 1\right\rangle )\left| \psi
\right\rangle .
\end{equation}
We have the $k$-separable state $\rho _{s0}=\int \left| \Omega
\right\rangle \left\langle \Omega \right| \prod_{i=1}^{N_1}\frac{d\varphi _i%
}{2\pi }$, which is
\begin{eqnarray}
\rho _{s0} &=&\frac 1{2^{N_1+1}}[2\left| GHZ_N\right\rangle \left\langle
\cdot \right|  \nonumber \\
&&+\sum_{\mathbf{j\neq 0;}j_1,\cdots ,j_{N_1}=0,}^1(\left| j_1\cdots
j_{N_1}0^{\otimes L}\right\rangle \left\langle \cdot \right|  \nonumber
\\
&&+\left| \overline{j_1}\cdots \overline{j_{N_1}}1^{\otimes
L}\right\rangle \left\langle \cdot \right| )],  \label{wee20}
\end{eqnarray}
where the binary string $\mathbf{j}=(j_1,j_2,\ldots ,j_{N_1}),$ $\overline{%
j_i}=1-j_i.$ Averaging over all the cases of qubit permutations we arrive at
a $k$-separable state
\begin{eqnarray}
\rho _{s1} &=&[2\left| GHZ_N\right\rangle \left\langle \cdot \right|
+N^{-1}N_1(T_1+T_{N-1})  \nonumber \\
&+&\sum_{i=2}^{N-2}(C_N^i)^{-1}(C_{N_1}^i+C_{N_1}^{i-L})T_i]/2^{N_1+1}.
\label{wee21}
\end{eqnarray}
where $T_i=\sum_{j_1+j_2+\cdots +j_N=i}\left| j_1j_2\ldots j_N\right\rangle
\left\langle \cdot \right| $. In convention, we have $C_{N_1}^i=0$ if $i>N_1$
and $C_{N_1}^{i-L}=0$ if $i<L.$

\begin{lemma}
The inequality
\begin{equation}
(C_{N_1}^i+C_{N_1}^{i-L})/C_N^i<N_1/N  \label{wee21a}
\end{equation}
is true for all $N$ and $N_1$ $(0<N_1<N-1).$
\end{lemma}

The proof is shown in appendix.

Denote
\begin{eqnarray}
\overline{\rho _{s2}} &=&%
\sum_{i=2}^{N-2}[N_1/N-(C_{N_1}^i+C_{N_1}^{i-N+N_1})/C_N^i]T_i/2^{N_1+1} \nonumber\\
&&+N^{-1}N_1(T_0+T_N)/2^{N_1+1},
\end{eqnarray}
where $T_0=\left| 0^{\otimes N}\right\rangle \left\langle \cdot \right| ,$ $%
T_N=\left| 1^{\otimes N}\right\rangle \left\langle \cdot \right| .$ Then $%
\overline{\rho _{s2}}$ is an unnormalized full separable state. We have $%
\overline{\rho _s}=\rho _{s1}+\overline{\rho _{s2}}$ being an unnormalized
$k$-separable ($k=N_{1}+1$) state in the sense that the system is divided into $N_1$ single
qubit parties and one $L$ qubit party. The normalized $k$-separable state is
\begin{equation}
\rho _s=\frac{\left| GHZ_N\right\rangle \left\langle \cdot \right| }{%
1+2^{N-1}N_1/N}+\frac{N_1/(2N)}{1+2^{N-1}N_1/N}I_{2^N}.  \label{wee22}
\end{equation}
Hence, for the partition $1^{N_1}|L$ (up to qubit permutation) the
noisy $N$ qubit GHZ state is separable iff
\begin{equation}
p\leq [1+2^{N-1}N_1/N]^{-1}.  \label{wee23}
\end{equation}

\subsection{Further splits of the last $L$ qubits}

We consider the partition $1^{N_1}|\overline{L},$ where $\overline{%
L}=L_1|L_2|\cdots |L_m$ is a split of $L$ with each $L_i$ containing
at least two qubits. Hence the number of parties is $k=N_{1}+m$. Suppose with respect to the partition $1^{N_1}|\overline{L}$ there is a $k$-separable pure state
\begin{equation}
\left| \Omega \right\rangle =\frac 1{\sqrt{2^{N_1+m}}}\prod_{i=1}^{N_1}(\left|
0\right\rangle +e^{i\varphi _i}\left| 1\right\rangle )\prod_{j=1}^{m}(\left|0^{\otimes L_{j}}
\right\rangle+ e^{i\varphi' _j}\left| 1^{\otimes L_{j}}\right\rangle ),
\end{equation}
with $\varphi' _m=-\sum_{i=1}^{N_1}\varphi _i-\sum_{j=1}^{m-1}\varphi _j$. We have the $k$-separable state $\rho _{s3}=\int \left| \Omega
\right\rangle \left\langle \Omega \right| \prod_{i=1}^{N_1}\frac{d\varphi _i%
}{2\pi }\prod_{j=1}^{m-1}\frac{d\varphi'_j%
}{2\pi }$, which is
\begin{eqnarray}\label{wee20}
\rho _{s3} &=&\frac 1{2^{N_1+m}}[2\left| GHZ_N\right\rangle \left\langle
\cdot \right|  \nonumber \\
&+&\sum_{\mathbf{j\neq 0}}(\left| j_1\cdots
j_{N_1}j_{N_1+1}^{\otimes L_{1}}\cdots j_{N_1+m-1}^{\otimes L_{m-1}}0^{\otimes L_{m}}\right\rangle \left\langle \cdot \right|  \nonumber
\\
&+&\left| \overline{j_1}\cdots \overline{j_{N_1}} \overline{j_{N_1+1}^{\otimes L_{1}}}\cdots \overline{j_{N_1+m-1}^{\otimes L_{m-1}}}1^{\otimes L_{m}}\right\rangle \left\langle \cdot \right| )],
\end{eqnarray}
Where the summation is over all $\mathbf{j}\in\{0,1\}^{\otimes (N_{1}+m)}$ except  $\mathbf{j= 0}$, and $\overline{j^{\otimes i}}=(1-j)^{\otimes i}$ for binary $j$.

\begin{lemma}
The averaging of $\rho_{s3}$ in (\ref{wee20}) over all the qubit permutations can be written as follows:
\begin{equation}\label{wee20a}
  \rho _{s4}=\frac 1{2^{N_1+m}}[2\left| GHZ_N\right\rangle \left\langle \cdot
\right| +\sum_{i=1}^{N-1}(C_N^i)^{-1}f_m(i)T_i],
\end{equation}
with
\begin{eqnarray}
   \sum_{i=1}^{N-1}f_m(i)=2^{N_1+m}-2,f_m(i)=f_m(N-i),f_m(1)\leq N \label{wee20b}\\
   0\leq f_m(i)\leq f_m(1)C_N^i/N  \text{\qquad for all   }  i\in [1,N-1]. \label{wee20c}
\end{eqnarray}
\end{lemma}
We will use mathematical induction to prove this lemma in the appendix.

 We then produce the noisy GHZ state by mixing some fully separable state with $\rho _{s4}$ due to conditions (\ref{wee20b}) and (\ref{wee20c}). So that the obtained $N$ qubit noisy GHZ state is $k$-separable with $k=N_{1}+m$.  We have the following sufficient criterion:

\begin{theorem}
The sufficient condition of $k$-separability ($k=N_{1}+m$) for a $N$ qubit noisy GHZ state is
\begin{equation}
p\leq[1+2^{N-1}N_1/N]^{-1}  \label{wee26}
\end{equation}
where $N_1>0$ is the number of parties with single qubit, the partition is $1^{N_{1}}|\overline{L}$ with $\overline{L}$ being $L_{1}|L_{2}|\cdot\cdot\cdot|L_{m}$, each partite $L_{j}$ has two or more qubits.
\end{theorem}

From (\ref{wee26}), we can see that the less the number $N_1,$ the larger
the right hand side of (\ref{wee26}), the better the sufficient condition. If the number of parties is
fixed to be $k$ ($k\geq \frac{N+1}2$)$,$ the smallest number of $N_1$ is $%
2k-N,$ we have the sufficient criterion:

\begin{theorem}
The sufficient condition of $k-$separability ($k\geq \frac{N+1}2$) for noisy
GHZ state is
\begin{equation}
p\leq [1+2^{N-1}(2k-N)/N]^{-1},  \label{wee27}
\end{equation}
The sufficient condition is realized by a proper state of the partition $%
1^{(2k-N)}|2^{(N-k)}$.
\end{theorem}

In figure 1, we display the $k$ partite separable conditions for $%
N-k=1,2,3,4,5.$
\begin{figure}[t]
\includegraphics[ trim=0.000000in 0.000000in -0.138042in 0.000000in,
height=2.5in, width=3.5in]{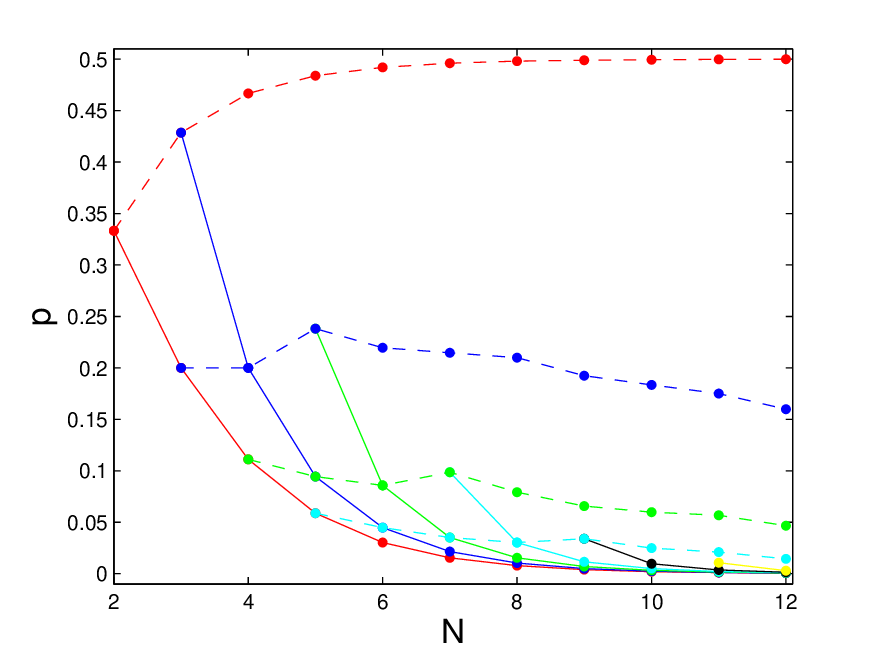}
\caption{(Color online)The entanglement properties of N-qubit GHZ states mixed with white
noise, $\rho _{ghzN}=p\left| GHZ_N\right\rangle \left\langle GHZ_N\right| +%
\frac{1-p}{2^N}I_{2^N}$. It was known before \protect\cite{Cirac},%
\protect\cite{Schack} that the states are fully separable iff $p\leq \frac
1{1+2^{N-1}}$ (the downmost solid line), The states biseparable iff $p\leq \frac{2^{N-1}-1}{2^N-1}$
\protect\cite{Seevinck}(the uppermost dashed line).  Our results show
that the states are $N-1$ separable iff $p\leq \frac 1{1+\frac{N-2}%
N2^{N-1}},$ $k$ separable iff $p\leq \frac 1{1+\frac{2k-N}N2^{N-1}}$ when $%
k\geq\frac{N+1}{2}$. They are shown by solid lines for $k$-separability with
$k=N,N-1,N-2,N-3,N-4,N-5$ from left to right. The dashed lines are for $k$-separability
with $k=2,3,4,5$ from top down.}
\end{figure}

\section{Other $k$-separability}

There are many $k$-separability noisy GHZ states that can not be fit into
the former regime of $k$ $(k\geq \frac{N+1}2)$ separability or
biseparability. The first case is the tri-separability of noisy GHZ states
for $N\geq 6.$ The partitions are $1|2|3$ and $2^{3}$ for $N=6.$ These
partitions give rise to the tripartite states:
\begin{equation}\label{40}
\rho _{1|2|3},\rho _{2^{3}}=\frac 1{2^3}(2\left| GHZ_6\right\rangle
\left\langle \cdot \right| +\sum_{i=1}^5v_i\frac{T_i}{C_6^i}),
\end{equation}
where $\mathbf{v}=(1,1,2,1,1)$ for $\rho _{1|2|3}$ and $\mathbf{v}%
=(0,3,0,3,0)$ for $\rho _{2^{3}}$, respectively. Consider the mixture of
this two kinds of partitions, the mixed state is $\rho _{s5}=$ $q\rho
_{1|2|3}+(1-q)\rho _{2^{3}}.$ We then have
\begin{equation}\label{41}
\rho _{s5}=\frac 1{2^3}(2\left| GHZ_6\right\rangle \left\langle \cdot
\right| +\sum_{i=1}^5u_i\frac{T_i}{C_6^i}),
\end{equation}

where $\mathbf{u}=(q,q+3(1-q),2q,q+3(1-q),q).$ Let $u_1/C_6^1=u_2/C_6^2\geq
u_3/C_6^3,$ then $q=\frac 23.$ Denote the coefficient of $T_1$ as $\frac
1\tau $, then $\tau =9.$ The tripartite separable sufficient condition is
\begin{equation}
p\leq \frac 1{1+2^5/\tau }=\frac 9{41}.  \label{wee42}
\end{equation}
To find the mixed state that has maximal $\tau $, we need to solve the
linear programming problem. This is because the mixture probability is
positive and we let $u_1/C_N^1=u_2/C_N^2=\cdots =u_i/C_N^i\geq
u_{i+j}/C_N^{i+j}$ to optimize $\tau $, for some $i\leq \left\lceil \frac
N2\right\rceil $ and all positive $j\leq \left\lceil \frac N2\right\rceil
-i. $

For $N\leq 12,$ we list all these mixed optimal states in Table I.

Table I. The $k$-separabilities of $N=6,...,12;$ $\tau \cdot fractions$ are $%
\tau $ times the fractions of partitions. For the tri-separable of $6$ qubit
noisy GHZ state, partitions $2^{3},1|2|3$ contribute $q=\frac 13$ and $%
1-q=\frac 23$ fractions in the optimization of $\tau .$ Hence $\tau \cdot
fractions=(3,6)$. $p_{s}$ is the critical value such that the noisy GHZ state is $k$-separable if $p\leq p_{s}$.

$
\begin{array}{llllll}
\hline\hline
N & k & partitions & \tau \cdot fractions & \tau & p_s \\ \hline
6 & 3 & 2^{3},1|2|3 & 3,6 & 9 & \frac 9{41} \\
7 & 3 & 2^{2}|3,1|3^{2} & 10.5,7 & 17.5 & \frac{35}{163} \\
8 & 3 & 1|3|4,2|3^{2},2^{2}|4 & 8,24,2 & 34 & \frac{17}{81} \\
8 & 4 & 1|2^{2}|3,2^{4} & 8,3 & 11 & \frac{11}{139} \\
9 & 3 & 1|4^{2},2|3|4,3^{3} & 16,36,9 & 61 & \frac{61}{327} \\
9 & 4 & 1|2|3^{2},2^{3}|3 & 9,9 & 18 & \frac 9{137} \\
10 & 3 & 1|4|5,2|4^{2},3^{2}|4 & 10,45,60 & 115 & \frac{115}{627} \\
10 & 4 & 1|3^{3},2^{2}|3^{2} & 10,22.5 & 32.5 & \frac{65}{1089} \\
10 & 5 & 1|2^{3}|3,2^{5} & 10,3 & 13 & \frac{13}{525} \\
11 & 3 &
\begin{array}{l}
3|4^{2},2|4|5, \\
3^{2}|5,1|5^{2}
\end{array}
&
\begin{array}{l}
137.5,55, \\
13.75,11
\end{array}
& \frac{869}4 & \frac{869}{4965} \\
11 & 4 &
\begin{array}{l}
2^{2}|3|4,2|3^{3}, \\
1|3^{2}|4
\end{array}
&
\begin{array}{l}
4.4,46.2, \\
11
\end{array}
& 61.6 & \frac{77}{1375} \\
11 & 5 & 2^{4}|3,1^{2}|3^{3} & 11,11 & 22 & \frac{11}{523} \\
12 & 3 &
\begin{array}{l}
4^{3},3|4|5, \\
2|5^{2},1|5|6
\end{array}
&
\begin{array}{l}
\frac{275}3,220, \\
66,12
\end{array}
& \frac{1169}3 & \frac{1169}{7313} \\
12 & 4 &
\begin{array}{l}
3^{4},2^{2}|4^{2}, \\
1|3|4^{2}
\end{array}
&
\begin{array}{l}
55,33, \\
12
\end{array}
& 100 & \frac{25}{537} \\
12 & 5 & 1|2|3^{3},2^{3}|3^{2} & 12,18 & 30 & \frac{15}{1039} \\
12 & 6 & 1|2^{4}|3,2^{6} & 12,3 & 15 & \frac{15}{2063} \\ \hline
\end{array}
$

\section{Discussion}

The $N$ qubit GHZ state as a codeword of the `tree' code was utilized to
explore the nonadditivity of channel capacity of depolarizing channel \cite
{Chen2011}. The `tree' code and the repetition code \cite{Smith} (or `cat'
code \cite{DiVincenzo}) have the same Hilbert subspace but with different
bases, they are the same code. A depolarizing channel $\mathcal{E}$ is a
completely positive trace preserving map, it maps an input qubit state $\rho
$ to an output state $\mathcal{E}(\rho )$ , with
\begin{equation}\label{42}
  \mathcal{E}(\rho )=(1-3q)\rho +q\sigma _1\rho \sigma _1+q\sigma _2\rho
\sigma _2+q\sigma _3\rho \sigma _3.
\end{equation}

Where the noise $q$ characterizes the depolarizing channel. When the $N$
qubit GHZ state passing through $N$ parallel depolarizing channels, the
output state is $\mathcal{E}^{\otimes N}(\left| GHZ_N\right\rangle
\left\langle GHZ_N\right| )$. The output state is a mixture of $N$ qubit GHZ
state and a color noise (a diagonal state in the computational basis), the
fraction of $\left| GHZ_N\right\rangle \left\langle GHZ_N\right| $ in the
output state is $(1-4q)^N$. We wonder if the output state is entangled,
partially separable or fully separable. We may use $\rho _{ghzN\text{ }}$ to
approximate $\mathcal{E}^{\otimes N}(\left| GHZ_N\right\rangle \left\langle
GHZ_N\right| ).$ In table II, we use $q_{th}$ to denote the threshold noise
(derived from $Fidelity=1-3q$ \cite{DiVincenzo}) of depolarizing channel.
Quantum coherent information is positive for the input of `tree' code
when the channel noise is less than the threshold noise, otherwise it is define to be zero. The codewords of the code
are $\left| GHZ_N\right\rangle $ and $\left| GHZ_{N-}\right\rangle =\frac 1{%
\sqrt{2}}(\left| 0^{\otimes N}\right\rangle -\left| 1^{\otimes
N}\right\rangle )$ and they have equal prior probabilities. We compare $%
q_{th}$ with the various critical values of $q$ for different separabilities
of $\rho _{ghzN\text{ }}.$ The critical values of $q$ are denoted as $%
q_{N-k} $ for the partition $1^{(2k-N)}|2^{(N-k)}.$ When $q>q_{N-k},$ the
state $\rho _{ghzN\text{ }}$ is separable for the partition $%
1^{(2k-N)}|2^{(N-k)}. $ When $q>q_0,$ the state $\rho _{ghzN\text{ }}$is
fully separable.

Notice that the biseparabilities of both $\rho _{ghzN\text{ }}$ and $%
\mathcal{E}^{\otimes N}(\left| GHZ_N\right\rangle \left\langle GHZ_N\right|
) $ are detected by the same witness $\frac{\Bbb{I}}2-\left|
GHZ_N\right\rangle \left\langle GHZ_N\right| ,$ they have the same
biseparable critical value $q_{bi}$ when they have the same fractions of $%
\left| GHZ_N\right\rangle .$ Table II shows us that $q_{th}>q_{bi}$ for all $%
N\geq 3.$ This means that the state $\mathcal{E}^{\otimes N}(\left|
GHZ_N\right\rangle \left\langle GHZ_N\right| )$ is entangled for $N=2$ and
not genuinely entangled for all $N\geq 3$ when the noise of channel arrives
the threshold.

On the other hand, $q_{th}<q_0.$ The noise of the output state $\mathcal{E}%
^{\otimes N}(\left| GHZ_N\right\rangle \left\langle GHZ_N\right| )$ is a
color noise. Thus some of the diagonal elements should be smaller than the
average noise level, which is the noise level of $\rho _{ghzN\text{ }}$ if
the same fractions of $\left| GHZ_N\right\rangle $ in the two states are
assumed. The smaller the diagonal element, the easier the entanglement takes
place. Hence, $\mathcal{E}^{\otimes N}(\left| GHZ_N\right\rangle
\left\langle GHZ_N\right| )$ is easier to be entangled than $\rho _{ghzN%
\text{ }},$ and tolerant a higher level of noise than the later to keep to
be entangled. Thus $q_{th}<q_0$ holds for output state $\mathcal{E}^{\otimes
N}(\left| GHZ_N\right\rangle \left\langle GHZ_N\right| ).$ So that when the state $\left|
GHZ_N\right\rangle $ is transmitted over depolarizing channel, the output state can not be fully separable if the channel noise is less than its threshold. The output state should maintain some kinds of entanglement.

Table II. Comparison of the threshold noise $q_{th}$ to various critical values of $q$ for different separabilities of $\rho _{ghzN%
\text{ }}.$

$
\begin{array}{lllllll}
\hline\hline
N & q_0 & q_1 & q_2 & q_3 & q_{th} & q_{bi} \\ \hline
2 & 0.1057 &  &  &  & 0.0628 & 0.1057 \\
3 & 0.1038 & 0.0615 &  &  & 0.0634 & 0.0615 \\
4 & 0.1057 & 0.0828 &  &  & 0.0633 & 0.0434 \\
5 & 0.1081 & 0.0941 & 0.0624 &  & 0.0635 & 0.0338 \\
6 & 0.1104 & 0.1010 & 0.0840 &  & 0.0634 & 0.0279 \\
7 & 0.1123 & 0.1056 & 0.0950 & 0.0704 & 0.0634 & 0.0238 \\ \hline
\end{array}
$

In summary, the output state of depolarizing channel with threshold noise is
neither genuinely entangled nor fully separable when we input GHZ state. It
is entangled for $N=2$ and partially entangled for $N\geq 3.$ Thus the $k$%
-separability makes sense.

\section{Conclusion}

We find the entanglement witnesses for almost a half of all kinds of entanglement of an $N$
qubit GHZ state in white noise. The witnesses are linear combinations of the
stabilizer group elements of GHZ states. The combinational coefficient of a
stabilizer element relies on the number of generators in the stabilizer
element. The necessary and sufficient condition for $k$ partite separability
has been given for arbitrary $N$ when $k\geq \frac{N+1}2$. The necessary and
sufficient condition is achieved by a partition (up to qubit permutations)
with $N-k$ double qubit parties and $2k-N$ single qubit parties. The $k$
separability condition is $p\leq [1+2^{N-1}(2k-N)/N]^{-1},$ where $p$ is the
fraction of pure GHZ state in the noisy $N$ qubit GHZ state. This gives rise
to the necessary and sufficient conditions for more than a half of the $k$%
-separability of multi-qubit noisy GHZ states. For all the other $k$%
-separable problems of noisy multi-qubit GHZ states, we find that the
sufficient separable conditions are achieved by the mixtures of the
partitions with the same $k$ (number of parties) but different qubit number
distributions. We display all the separabilities of $N$ qubit noisy GHZ
states for $N\leq 12.$ As an application, we have approximated the output of
GHZ state passing through depolarizing channel with the GHZ state in white
noise. It has been shown the output state is neither genuinely entangled nor
fully separable when the depolarizing channel has threshold noise.

GHZ states are the simplest graph states. The method of constructing
entanglement witness operators with linear combinations of stabilizer
operators in this paper is also useful for entanglement detection of noisy
graph states. A stabilizer codeword passing through certain quantum channel
may become a noisy graph state. Thus our method may find more applications
in quantum information transmission.

\section*{Acknowledgment}

Supported by the National Natural Science Foundation of China (Grant Nos.
11375152) and (partially) supported by National Basic Research Program of
China (Grant No. 2014CB921203) are gratefully acknowledged.

\section*{Appendix}

\subsection*{Proof of Lemma 1}

Proof: In computational basis, we use $\left| \psi \right\rangle =\sum_{%
\mathbf{i}}\alpha _{\mathbf{i}}\left| \mathbf{i}\right\rangle $ to denote
the pure state of the last $L$ qubits in $\rho _s$ with the partition $%
1^{N_1}|L.$ Where $\mathbf{i=(}i_1,i_2,\ldots ,i_{L})$
is a binary string and we denote the weight of $\mathbf{i}$\textbf{\ }as $%
\left| \mathbf{i}\right| $. We rewrite (\ref{wqq1}) as
\begin{equation}
tr(\rho _s\hat{M}_{D})=\sum_{i=1}^{\left\lfloor N/2\right\rfloor
}M_i\sum_{n=\max \{0,2i-N_1\}}^{\min \{L,2i\}}S_{2i-n}q_n.  \label{wee28}
\end{equation}
Where $S_0=1,S_i=\sum_{1\leq j_1<j_2<\ldots <j_i\leq
N_1}\prod_{l=1}^iz_{j_l},$and
\begin{eqnarray}
q_n &=&\sum_{\left| \overline{\mathbf{j}}\right| =n}\left\langle \psi
\right| \bigotimes_{l=N_1+1}^N\sigma _3^{j_l}\left| \psi \right\rangle \nonumber \\
&=&\sum_{\left| \overline{\mathbf{j}}\right| =n}\sum_{\mathbf{i}}\left|
\alpha _{\mathbf{i}}\right| ^2(-1)^{\mathbf{i\cdot }\overline{\mathbf{j}}},
\end{eqnarray}
where $\overline{\mathbf{j}}=(j_{N_1+1},\cdots ,j_N)$ is a binary string
and $\left| \overline{\mathbf{j}}\right| $ is the weights of $\overline{%
\mathbf{j}}$. We may write $q_n=\sum_{l=0}^{L}\sum_{\left| \mathbf{i}%
\right| =l}\left| \alpha _{\mathbf{i}}\right| ^2w_{n,l}$ with
\begin{equation}
w_{n,l}=\sum_{j=\max \{0,n+l-L\}}^{\min
\{n,l\}}(-1)^jC_{L-l}^{n-j}C_l^j.  \label{wee29}
\end{equation}
Then
\begin{equation}
tr(\rho _s\hat{M}_{D})=\sum_{l=0}^{L}\sum_{\left| \mathbf{i}\right|
=l}\left| \alpha _{\mathbf{i}}\right| ^2\Gamma _l,  \label{wee29s}
\end{equation}
with
\begin{eqnarray}
\Gamma _l &=&\sum_{m=1}^{\left\lceil \frac{N_1}2\right\rceil
}S_{2m-1}\sum_{i=1}^{\left\lceil \frac{L}2\right\rceil
}M_{i+m-1}w_{2i-1,l}  \nonumber \\
&&+\sum_{m=0}^{\left\lfloor \frac{N_1}2\right\rfloor
}S_{2m}\sum_{i=0}^{\left\lfloor \frac{L}2\right\rfloor }M_{i+m}w_{2i,l}.
\label{wee29a}
\end{eqnarray}

If we denote
\begin{eqnarray}
\mathcal{X} &=&\sum_{\left| \overline{\mathbf{j}}\right| |even}(-1)^{\left|
\overline{\mathbf{j}}\right| /2}\bigotimes_{l=N_1+1}^N\sigma
_1^{1-j_l}\sigma _2^{j_l}, \\
\mathcal{Y} &=&\sum_{\left| \overline{\mathbf{j}}\right| |odd}(-1)^{(\left|
\overline{\mathbf{j}}\right| -1)/2}\bigotimes_{l=N_1+1}^N\sigma
_1^{1-j_l}\sigma _2^{j_l},
\end{eqnarray}
Equation (\ref{wqq2}) reads
\begin{equation}
tr(\rho _s\hat{M}_{A})=\left\langle \psi \right| c\mathcal{X}-d\mathcal{Y}%
\left| \psi \right\rangle ,  \label{wee30}
\end{equation}
Then we have
\begin{eqnarray}
\left\langle \psi \right| \mathcal{X}\left| \psi \right\rangle
&=&2^{L-1}(\alpha _{\mathbf{0}}\alpha _{\mathbf{1}}^{*}+\alpha _{\mathbf{%
0}}^{*}\alpha _{\mathbf{1}}),  \label{wee31} \\
\left\langle \psi \right| \mathcal{Y}\left| \psi \right\rangle
&=&i2^{L-1}(\alpha _{\mathbf{0}}\alpha _{\mathbf{1}}^{*}-\alpha _{%
\mathbf{0}}^{*}\alpha _{\mathbf{1}}),  \label{wee32}
\end{eqnarray}
We have used $\left\langle 0^{\otimes (L-l)}1^{\otimes l}\right|
\mathcal{X}\left| 1^{\otimes (L-l)}0^{\otimes l}\right\rangle =0$ if $%
l\neq 0$ and $l\neq L.$ This is due to the fact that for each term in $%
\mathcal{X}$ if there are odd number of $\sigma _2$ in the last $l$ qubits,
then there are odd number of $\sigma _2$ in the first $L-l$ qubits too,
so that an extra $-1$ factor emerges due to $\sigma _2^2$ for such a term.
The probabilities of odd and even number of $\sigma _2$ in the last $l$
qubits are equal. The terms of $\mathcal{X}$ are concealed with each other
in the evaluation of the matrix element $\left\langle 0^{\otimes
(L-l)}1^{\otimes l}\right| \mathcal{X}\left| 1^{\otimes
(L-l)}0^{\otimes l}\right\rangle $. The null result is also true for $%
\mathcal{Y}$ when $l\neq 0$ and $l\neq L.$

From (\ref{wee29}), we have $w_{i,0}=C_{L}^i,$ $%
w_{i,L}=(-1)^iC_{L}^i$. Thus
\begin{eqnarray}
\Gamma _0 &=&N/N_1+2^{L-1}(a+b),  \label{wqq3} \\
\Gamma _{L} &=&N/N_1+2^{L-1}(a-b),  \label{wqq4}
\end{eqnarray}
where
\begin{eqnarray}
a &=&-1+N_1{}^{-1}\sum_{m=1}^{\left\lfloor N_1/2\right\rfloor
}[4m-N_1]S_{2m}, \\
b &=&N_1{}^{-1}\sum_{m=1}^{\left\lceil N_1/2\right\rceil }[4m-2-N_1]S_{2m-1}.
\end{eqnarray}
$\blacksquare $

\subsection*{Proof of Lemma 2}

Proof: This is equivalent to $a+\sqrt{b^2+c^2+d^2}\leq 0$. We should show
that (i) $a\leq 0,$ (ii) $a^2-b^2\geq c^2+d^2.$ Let
\begin{eqnarray}
u &=&N_1{}^{-1}\sum_{j=1}^{N_1}\prod_{n=1}^{N_1}[1+(-1)^{\delta _{j,n}}z_n],
\\
v &=&N_1{}^{-1}\sum_{j=1}^{N_1}\prod_{n=1}^{N_1}[1-(-1)^{\delta _{j,n}}z_n].
\end{eqnarray}
A direct calculation shows that
\begin{eqnarray}
u &=&1+\sum_{i=1}^{N_1}[1-2i/N_1]S_i, \\
v &=&1+\sum_{i=1}^{N_1}[1-2i/N_1](-1)^iS_i.
\end{eqnarray}
Hence $\frac 12(u+v)=-a$ and $\frac 12(u-v)=-b$ . We arrive at (i) $a\leq 0,$
since $u\geq 0$ and $v\geq 0$ due to $z_n\in [-1,1]$ for all $n.$ We further
have $a^2-b^2=uv,$ they are
\begin{equation}
N_1{}^{-2}\sum_{i=1}^{N_1}\sum_{j=1}^{N_1}(1-z_i)^2(1+z_j)^2\prod_{n=1,n\neq
i,j}^{N_1}(1-z_n^2),
\end{equation}
meanwhile $c^2+d^2=$ $\prod_{i=1}^{N_1}(1-z_i^2),$ thus we arrive at (ii)
\begin{eqnarray}
a^2-b^2-c^2-d^2 &=&8N_1{}^{-2}\sum_{1\leq i<j\leq N_1}(z_i-z_j)^2 \\
\times \prod_{n=1,n\neq i,j}^{N_1}(1-z_n^2) &\geq &0.
\end{eqnarray}
$\blacksquare $

\subsection*{Proof of Theorem 1}

Proof: The matrix $\mathcal{M}=\bigoplus_{\mathbf{i\neq 0,1}}\mathcal{M}_{%
\mathbf{i,i}}\oplus \mathcal{M}_0$. Except for the submatrix $\mathcal{M}_0,$
the other eigenvalues of $\mathcal{M}$ are just the other diagonal elements $%
\mathcal{M}_{\mathbf{i,i}}=\Gamma _{\left| \mathbf{i}\right| }$. From the
definition of $w_{n,l}$ in (\ref{wee29}), it is not difficult to show that
\begin{eqnarray}
\sum_{i=0}^{\left\lfloor \frac{L}2\right\rfloor }w_{2i,l}
&=&0,\sum_{i=1}^{\left\lceil \frac{L}2\right\rceil }w_{2i-1,l}=0,
\label{wee37} \\
\sum_{i=0}^{\left\lfloor \frac{L}2\right\rfloor }iw_{2i,l} &=&\left\{
\begin{array}{l}
0,\text{ for }1<l<L-1 \\
-2^{L-3},\text{ for }l=1,L-1
\end{array}
\right.  \label{wee38} \\
\sum_{i=1}^{\left\lceil \frac{L}2\right\rceil }iw_{2i-1,l} &=&\left\{
\begin{array}{l}
0,\text{ for }1<l<L-1 \\
-2^{L-3},\text{ for }l=1 \\
2^{L-3},\text{ for }l=L-1
\end{array}
\right.  \label{wee39}
\end{eqnarray}
Using (\ref{wee34}), we have
\begin{eqnarray}
\sum_{i=1}^{\left\lceil \frac{L}2\right\rceil }M_{i+m-1}w_{2i-1,l}=\frac{%
2^{L-1}}{N_1}(-\delta _{l,1}+\delta _{l,L-1}),\\
\sum_{i=0}^{\left\lfloor \frac{L}2\right\rfloor }M_{i+m}w_{2i,l}=\frac
N{N_1}\delta _{m,0}-\frac{2^{L-1}}{N_1}(\delta _{l,1}+\delta
_{l,L-1}).
\end{eqnarray}
From (\ref{wee39}), for $1<l<L-1,$ we have
\begin{equation}
\Gamma _l=\frac N{N_1};  \label{wqq5}
\end{equation}
For $l=1,$ we arrive at
\begin{eqnarray}
\Gamma _1 &=&\frac N{N_1}-\frac{2^{L-1}}{N_1}\sum_{m=0}^{N_1}S_m \nonumber\\
&=&\frac N{N_1}-\frac{2^{L-1}}{N_1}\prod_{m=1}^{N_1}(1+z_m) \nonumber \\
&\leq &\frac N{N_1},
\end{eqnarray}
and
\begin{eqnarray}
\Gamma _{L-1} &=&\frac N{N_1}-\frac{2^{L-1}}{N_1}%
\prod_{m=1}^{N_1}(1-z_m) \nonumber\\
&\leq &\frac N{N_1}.
\end{eqnarray}
Thus we have proved that all the eigenvalues of $\mathcal{M}$ are up bounded
by $\frac N{N_1}.$ The up bound is achievable. Hence for our choice of the
witness in (\ref{wee34}), the largest eigenvalue of the matrix $\mathcal{M}$%
, thus the maximum of $tr(\rho _s\hat{M})$ is $\frac N{N_1}$ for the
partition $1^{N_1}|L.$ The $M_i$ in (\ref{wee34}) yields $%
\sum_{i=1}^{\left\lfloor N/2\right\rfloor }M_iC_N^{2i}=\frac N{N_1}.$ $%
\blacksquare $

\subsection*{Proof of Lemma 3}

Proof: For $i>N_1$ or $i<L$, the inequality (\ref{wee21a}) is apparently
true. So we consider $L\leq i\leq N_1.$ Since $L>1$ is assumed, we
have $1<i<N-1.$ The inequality can be rewritten as
\begin{eqnarray}
&&(N-i)(N-i-1)\cdots (N-i-L+1) \nonumber\\
&&+i(i-1)\cdots (i-L+1) \nonumber\\
&<&(N-1)(N-2)\cdots N_1.
\end{eqnarray}
For $L=2,3,$ the inequality becomes $N>i+1,$ which is true. Let the
inequality be true for $L=l,$ we will show that the inequality will be
true for $L=l+1$ too. Notice that $N>i+1,$ we have $\frac{i-l}{N-l-1}<1.$
Meanwhile $i>1$ leads to $\frac{N-i-l}{N-l-1}<1.$ Thus we have
\begin{eqnarray}
&&(N-i)(N-i-1)\cdots (N-i-l+1)\frac{N-i-l}{N-l-1} \nonumber\\
&&+i(i-1)\cdots (i-l+1)\frac{i-l}{N-l-1} \nonumber\\
&<&(N-1)(N-2)\cdots (N-l),
\end{eqnarray}
which is the inequality for $L=l+1.$ $\blacksquare $

\subsection*{Proof of Lemma 4}
When $m=1$, the partition reduces to $1^{N_1}|L$, the state $\rho _{s4}$ reduces to $\rho _{s1}$. We can verify that (\ref{wee20b}) and (\ref{wee20c}) are true for $m=1$ case.

Let equation (\ref{wee20b}) be true for $m=m'$ case.
For the case of $m=m'+1$, let us consider a partition $1^{N_1}|\overline{L}|l$ of $N+l$ qubit system with $l\geq 2$. The state for this partition can be described by
\begin{equation}\label{wee20e}
  f_{m'+1}(i)=\left\{
\begin{array}{l}
f_{m'}(i)+f_{m'}(i-l)\text{ for }i\in [l,N]; \\
f_{m'}(i)\text{ for }i<l; \\
f_{m'}(i-l)\text{ for }i>N.
\end{array}
\right.
\end{equation}
We want show that $f_{m'+1}(i)/C_{N+l}^i\leq f_{m'+1}(1)/(N+l)$ for $i\in [1,N+l-1].$
Apparently, we only need to show it for $i\in [l,N].$ We have $%
f_{m'+1}(i)=f_{m'}(i)+f_{m'}(i-l)\leq f_{m'}(1)(C_N^i+C_N^{i-l})/N.$ We have $f_{m'+1}(1)=f_{m'}(1)
$ for $l\geq 2$ (in fact $f_{m'}(1)=N_1$)$,$ thus what we need to show is
\begin{equation}
(C_N^i+C_N^{i-l})/N\leq C_{N+l}^i/(N+l).
\end{equation}
It is the inequality proved in Lemma 3. Thus for the partition $1^{N_1}|%
\overline{L}|l$ of the $N+l$ qubit noisy GHZ state with $l\geq 2,$ we
have $0\leq f_{m'+1}(i)\leq f_{m'+1}(1)C_{N+l}^i/(N+l)$ for all $i\in [1,N+l-1].$
Hence the assumption (\ref{wee20a}) is true. $%
\blacksquare $

\end{document}